*Perspective*

# Autferroicity: concept, candidates, and applications

Jun-Jie Zhang, Ziwen Wang, and Shuai Dong*

Key Laboratory of Quantum Materials and Devices of Ministry of Education, School of Physics, Southeast University, Nanjing 211189, China

*Corresponding author (email: sdong@seu.edu.cn)

Autferroicity is a newly proposed form of hybrid ferroicity, which is a sister branch of multiferroicity. It is characterized by the mutually exclusive magnetic and polar phases within a single system, giving a unique seesaw-type magnetoelectric coupling. This perspective provides a theoretical overview of its underlying concept, phase diagram characteristics, and representative candidates such as Ti-based trichalcogenide monolayers, while also highlighting its potential applications in nonvolatile memory devices and true random number generation.

Magnetism and electric polarity are two fundamental physical phenomena in solids which can be coupled, giving rise to the magnetoelectric (ME) coupling. The concept of magnetoelectricity currently extends beyond its original definition, representing all forms of interactions between spin and charge degrees of freedom [1,2,3]. As defined by H. Schmid in 1994 [4], multiferroics are a class of materials in which both two or more primary ferroic properties (ferroelectricity, ferromagnetism, ferroelasticity) simultaneously coexist within the same phase. The coexisting of magnetism and electric polarity provide the opportunity to realize cross-control mechanisms [5], making multiferroics highly attractive for applications in multifunctional devices [6].

However, ferroelectricity and magnetism often work against each other, making it challenging for them to coexist in a single material. In 2000, N. Hill provided a theoretical explanation for the difficulty in the discovery of new multiferroics [7]. For example, spontaneous polarizations in perovskite ferroelectrics arise from non-centrosymmetric structural distortions, e.g. as in $BaTiO_3$, which typically require empty $d$ orbitals to flexible bonding and hybridization with surrounding anions, i.e., the so-called $d^0$ rule. In contrast, magnetic moments in magnetic materials originate from unpaired electrons in partially filled $d$ and/or $f$ orbitals. This fundamental incompatibility makes the coexistence of ferroelectricity and magnetism uncommon in solids, and thus multiferroics are highly nontrivial.

The discovery of multiferroicity in hexagonal YMnO$_3$ ($h$-YMnO$_3$) represents an early and typical example in which the mutual exclusion between ferroic orders is circumvented through unconventional mechanisms [8,9]. The ferroelectricity in $h$-YMnO$_3$ is caused by geometric lattice distortions, thereby avoiding the constraint of $d^0$ rule and allowing compatibility with magnetism from partially filled $d$ orbitals. Since the primary order parameter of its ferroelectricity is not the electric polarization itself, $h$-YMnO$_3$ is classified as an improper ferroelectric material [10]. Other types of multiferroics with improper ferroelectricity include charge-order-driven (e.g., LuFe$_2$O$_4$, Ca$_3$Ru$_2$O$_7$, and LiFe$_2$F$_6$) [11,12,13] and magnetically-induced (e.g., TbMnO$_3$ and TbMn$_2$O$_5$) electric polarity [14,15]. Hence, improper ferroelectricity has become a key concept in the search for novel single-phase multiferroics with strong ME coupling.

Nevertheless, single-phase multiferroics can also incorporate proper ferroelectricity in some special cases, e.g., BiFeO$_3$ [16]. In BiFeO$_3$, magnetism and electric polarization arise from different sources: the $6s^2$ lone pair of Bi$^{3+}$ ion gives rise to ferroelectric dipole, while the partially filled Fe$^{3+}$'s $3d^5$ orbitals are responsible for its magnetic moment. Despite the different origins of the ferroicities in BiFeO$_3$, its ME coupling is mainly governed by the Dzyaloshinskii-Moriya interaction [17].

**Limits of multiferroics**

While the mutual exclusion between magnetism and electric polarity can be circumvented in multiferroics, trade-offs between ferroic orders and their ME coupling still persevere. Specifically, the polarizations in multiferroics with improper ferroelectricity are typically much weaker than those in proper ferroelectrics [2,18], as they arise as high-order effects. The typical values of improper ferroelectricity are below 10 μC/cm$^2$, in contrast to ~100 μC/cm$^2$ for BiFeO$_3$ with proper ferroelectricity [16]. In terms of ME coupling, those type-I multiferroics, e.g. BiFeO$_3$, exhibit relatively weak coupling despite their strong ferroelectricity, owing to the independent origins of ferroelectric and magnetic orders [16]. In 2009, Khomskii categorized mutiferroics into two main types [19]: type-I multiferroics have independent ferroic origins, while type-II exhibit magnetically driven ferroelectricity with strong ME coupling. In contrast, the type-II multiferroics, such as TbMnO$_3$, display much stronger ME coupling because their electric polarizations are directly induced by magnetic orders [14]. However, this comes at the expense of weaker ferroelectricity, almost the lowest in improper ferroelectrics. Hence, such trade-offs appear to be unavoidable and are unlikely to be fully resolved within the conventional framework of multiferroics.

On the other hand, two-dimensional (2D) materials have attracted significant attentions since the successful mechanical exfoliation of graphene from graphite [20]. More recently, a growing number of 2D multiferroics have been theoretically predicted, revealing several new mechanisms for achieving ME coupling [21,22,23]. For instance, the realization of proper ferroelectricity in 2D systems is no longer restricted to the conventional $d^0$ rule or the $6s$ lone pairs. Moreover, 2D type-II multiferroicity has been predicted in MXenes and $VX_2$ (X = Cl, Br, I) monolayers [24,25], exhibiting dominant electronic contributions to polarization (e.g., ~98.5% in $Hf_2VC_2F_2$) [24], and has been experimentally confirmed in $NiI_2$ monolayer [26]. Furthermore, the concept of a 2D hyper-ferroelectric metal has been proposed [27]. Very recently, Jiang *et al.* proposed the concept of type-III multiferroicity in a metal halide monolayer [28], in which magnetism is driven by ferroelectricity, enabling effective electrical control of magnetic order. Despite these advances, strong ME coupling identified in reported 2D multiferroics is still fundamentally based on high-order effects. Consequently, the realization of multiferroics that simultaneously exhibit robust ferroic orders and strong ME coupling remains challenging for practical implementation.

**Concepts of autferroicity**

To overcome the challenge in multiferroics, recently we proposed a new form of hybrid ferroicity as a sister branch of multiferroicity [29,30], which was initially termed ***alterferroicity*** [alter(native)+ferroicity] [29,30]. However, to avoid the confusion with the emerging field of altermagnetism [alter(nating)+magnetism], we rename it as ***autferroicity***, with the prefix *aut-* derived from Latin, meaning "or" or "either" [30].

In autferroicity, the coexistence of magnetism and electric polarity in the same phase is no longer expected; instead, their natural mutual exclusion is utilized as a functional advantage. Crucially, autferroicity exhibits the intrinsically strong ME coupling, i.e., the so-called seesaw-type magnetoelectricity (Fig. 1a) [29], which is conceptually different from both conventional linear and nonlinear ME effects in multiferroics. With this seesaw-type ME coupling, the balance between magnetic and polar phases can be tuned within the framework of their mutual exclusivity, enabling the switching of magnetism (or polarization) on and off via applied electric (or magnetic) fields in an autferroic material.

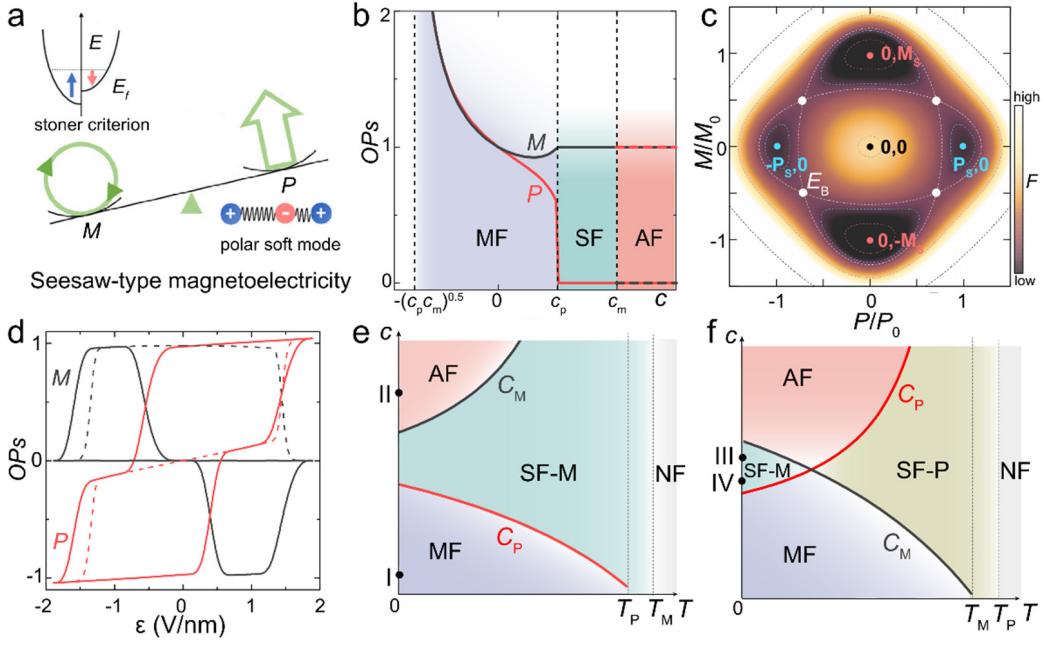

**Figure 1** (a) Schematic illustration of the seesaw-type magnetoelectricity in autferroics: the competition between electronic instability (left) and phononic instability (right). Reprinted with permission from Ref. [29]. Copyright 2023, the Authors, licensed under CC BY 4.0. (b) Ground state phase diagram as a function of the ME coupling coefficient $c$. MF: type-I multiferroic phase; SF: single-ferroic phase (magnetic for $c_m > c_p$); AF: autferroic phase. Order parameters (*OPs*) are shown as curves; dashed lines in the autferroic region indicate two alternative solutions. (c) Representative ($P$, $M$) energy landscape in the autferroic regime. $E_B$: energy barrier between two ferroic phases. (d) Hysteresis loop of autferroic under electric field, illustrating field-tunable switching between magnetic and polar states. (e-f) Analytical finite-temperature ($T$) phase diagrams in the $c$-$T$ parameter space. NF: nonferroic state; SF-M (SF-P): single-ferroic magnetic (ferroelectric) phase. Phase boundaries correspond to $C_P(T^*_P)$ (red) and $C_M(T^*_M)$ (black). $C_M$ and $C_P$ indicate finite-temperature transition behaviors of parameters $c_m \equiv 2bd/a$ and $c_p \equiv 2ae/d$, respectively. (e) $T_M > T_P$; (f) $T_P > T_M$. (b-f) Reprinted with permission from Ref. [30]. Copyright 2025, American Physical Society.

The Landau theory offers a systematic framework for describing autferroicity, emphasizing the qualitative rotation in the free energy landscape due to strong seesaw-type ME coupling. The canonical Landau free-energy model is expressed as [31,32]:

$$F(P,M,T) = \left[ -a\left(1 - \frac{T}{T_P}\right)P^2 + bP^4 \right] + \left[ -d\left(1 - \frac{T}{T_M}\right)M^2 + eM^4 \right] + cP^2M^2 \quad (1)$$

where $P$ and $M$ are order parameters for ferroelectricity and ferromagnetism, respectively. $T_P$ and $T_M$ are their respective transition temperatures; and coefficients $a,b$ ($d,e$) are Landau parameters for ferroelectricity (ferromagnetism), determining the corresponding ferroic transition behavior. Symmetry constraints imply that the lowest-order term for seesaw-type ME coupling takes the form $cP^2M^2$, where $c > 0$. Notably, this form of coupling is generally applicable to type-I multiferroics [1], independent of specific crystal symmetry. Note that the Eq. 1 can be also applied to antiferromagnetic systems, just by replacing $M$ using the antiferromagnetic order parameter $L$.

Unlike those type-I multiferroics, which typically yields weak biquadratic ME interactions, the defining characteristic of autferroicity is a large and positive coupling coefficient, representing a robust mutual exclusivity between the ferroic orders [30]. Specifically, the system favors a multiferroic phase when the positive coupling strength $c$ is smaller than both critical values (Fig. 1b), $c_m \equiv 2bd/a$ and $c_p \equiv 2ae/d$. A single-ferroic phase is stabilized when $c$ lies between these two thresholds (i.e., $c_m > c > c_p$ or $c_p > c > c_m$) (Fig. 1b), wherein one ferroic order defines the ground state while the other remains energetically unfavorable, occupying a saddle point of energy landscape. Remarkably, autferroicity emerges when the $c$ exceeds both $c_p$ and $c_m$ (Fig. 1b), giving rise to an energy landscape in which magnetic and electric states are individually stable yet mutually exclusive, separated by distinct energy barriers (Fig. 1c). Although the ME coupling of autferroics remains at the fourth-order level in the Landau free energy as in the type-I multiferroics, the much larger ME coefficient $c$ enables the complete and reversible switch of the system between magnetic and ferroelectric states (Fig. 1d), which is beyond the capacity of type-I multiferroics.

Autferroic transitions exhibit complex behavior at finite temperatures [30]. When the transition temperatures ($T_P$ and $T_M$) and energy depths ($F_P$ and $F_M$) of the two ferroic orders are symmetric (Fig. 1e), e.g., $T_P > T_M$ and $F_P > F_M$, the system typically transitions first into a single-ferroic state upon heating. The autferroic transition temperature is governed not only by the energy barrier separating the magnetic and ferroelectric states, but also by the difference between their intrinsic transition temperatures. A larger energy barrier or smaller $|T_P - T_M|$ tends to stabilize the autferroic phase over a broader temperature range. This behavior reflects the intrinsic seesaw nature of autferroicity and defines the temperature window where each ferroic state can be stabilized. Another interesting case arises when the transition temperatures and energy depths of the two ferroic orders are asymmetric (Fig. 1f), e.g., $T_P > T_M$ but $F_P < F_M$. In such situations, the system may transition from a single-ferroic state into either an autferroic or multiferroic phase within a finite temperature window, sensitively depending on the ME coupling strength (Fig. 1f).

**Candidates for autferroics**

In autferroics, the ferroic orders do not coexist simultaneously within the same phase. Accordingly, the primary design criterion for realizing autferroicity is the identification of materials exhibiting dual instabilities, specifically, a dynamical lattice instability favoring ferroelectric distortion and an electronic structure instability supporting magnetism (i.e. the so-called Stoner criterion). The second key criterion involves ensuring the mutual exclusivity of these two instabilities. In particular, it is better to have variable valences for the key ion, which can stabilize magnetic and ferroelectric phases respectively. This condition is often met in materials containing redox-active elements such as $Ti^{3+}/Ti^{4+}$ or $Cu^{1+}/Cu^{2+}$. Moreover, achieving a high autferroic phase transition temperature requires that the strengths of the dual instabilities be comparable. Thus, materials situated near the phase boundary between magnetic and polar ground states are particularly well-suited for autferroicity. In practice, a critical design principle is to target systems with partially covalent bonding between metal cations and surrounding anions, as the resulting "soft" chemical valences allow for fine tuning of the balance between polarity and magnetism, an advantage condition for stabilizing autferroic behavior.

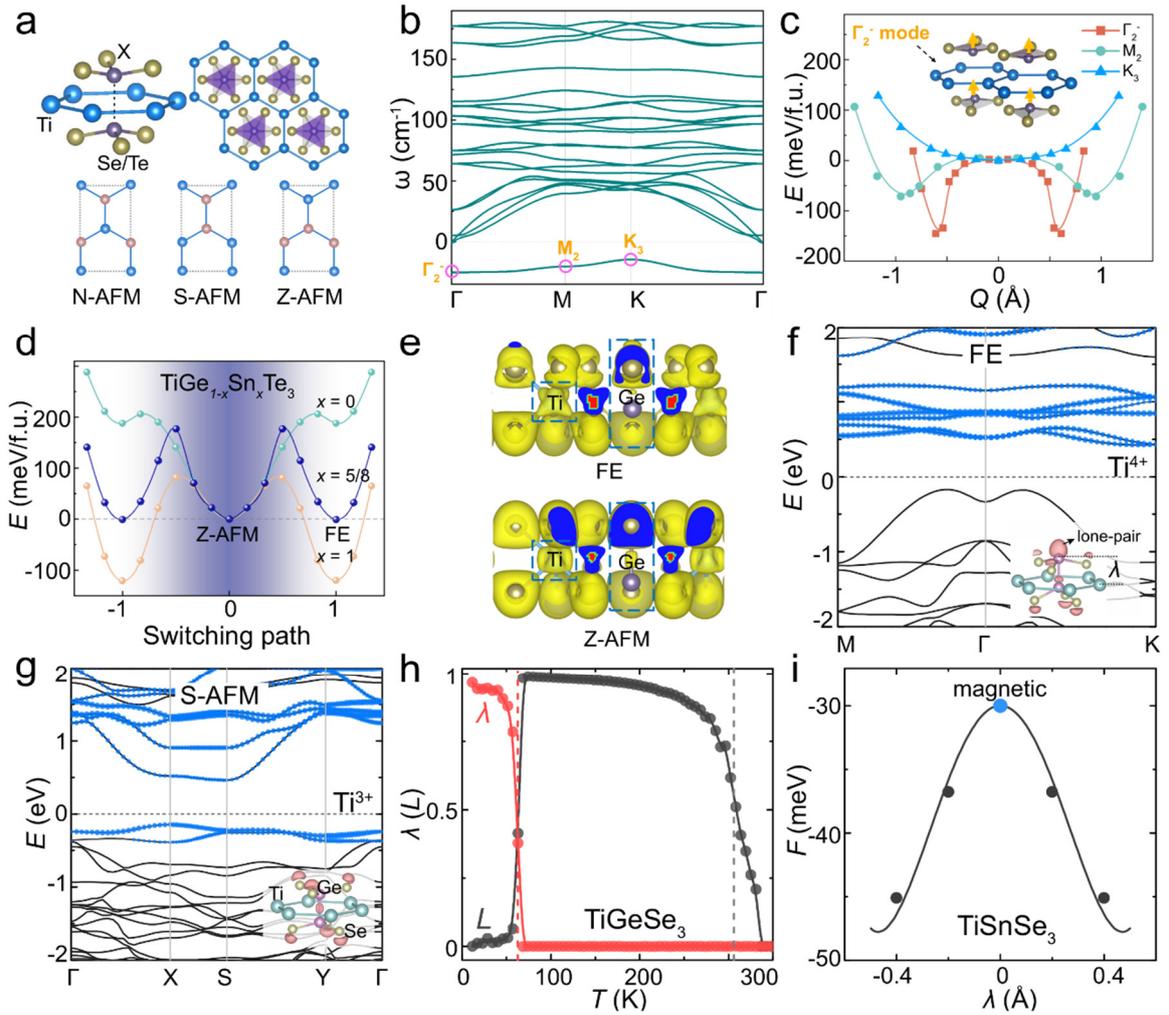

**Figure 2** Autferroic candidates based on Ti-based trichalcogenide monolayers. (a) Crystal

structures of Ti$XC_3$ monolayers ($X$=Ge, Sn; $C$=Se, Te), along with three possible antiferromagnetic configurations: Néel-type (N-AFM), stripy-type (S-AFM), and zigzag-type (Z-AFM). (b) Phonon instability in the high-symmetry nonmagnetic phase (space group $P$-31$m$) of TiGeTe$_3$. (c) Energy profiles of structural distortion modes in the absence of magnetism; the $\Gamma_2^-$ mode is energetically favored. Inset: schematic of the polar atomic displacement along the $c$-axis associated with the $\Gamma_2^-$ mode. (d) Energy profile along switching paths in autferroic Ti(Ge$_{1-x}$Sn$_x$)Te$_3$ monolayers. (e) Comparison of valence electron distribution of TiGeTe$_3$ for ferroelectric (FE) and Z-AFM phase. (a-e) Reprinted with permission from Ref. [29]. Copyright 2023, the Authors, licensed under CC BY 4.0. (f-g) Comparison of band structures between the (f) FE states and (g) S-AFM in TiGeSe$_3$. Blue bands: Ti's 3$d$ orbital projections. (h) Finite-temperature behavior of TiGeSe$_3$: evolution of the polar displacement amplitude ($\lambda$) and antiferromagnetic order parameter $L=(M_\uparrow - M_\downarrow)/2$ vs. temperature $T$. The initial ferroelectricity vanishes at ~64 K. (i) DFT evidence for the single-ferroic ground state in TiSnSe$_3$. (f-i) Reprinted with permission from Ref. [30]. Copyright 2025, American Physical Society.

Transition metal trichalcogenide (TMTC) monolayers represent a compelling class of candidate materials for autferroicity [29]. TMTC's constitute a large family of van der Waals layered crystals with the general chemical formula $ABC_3$ (Fig. 2a) [33], where $A$ is a transition metal, $B$ is an auxiliary element, typically a pnictogen or a group IV element, and $C$ is a chalcogen other than oxygen. This material class exhibits considerable structural and chemical diversity, accommodating both magnetic and polar phases. Typical examples include the ferroelectric CuInP$_2$S$_6$ [34] and the ferromagnetic CrGeTe$_3$ monolayer [35]. More importantly, the low electronegativities of the chalcogen and auxiliary ions promote partial covalent bonding, which enables valence flexibility.

Among the TMTC family, TiGeTe$_3$ monolayer was predicted as the first candidate material for autferroicity [29], based on first-principles density functional theory (DFT) calculations. In TiGeTe$_3$, Ti ions form a honeycomb lattice, with each Ti atom coordinated by six Te atoms in a TiTe$_6$ triangular antiprism geometry (Fig. 2a). According to DFT results, the Ge-Ge pairs undergo out-of-plane displacements driven by a softened polar phonon mode (Fig. 2b), resulting in a structural transition from a nonpolar to a purely ferroelectric phase (Fig. 2c).

In addition to its predicted ferroelectricity, we also identified that the nominal Ti$^{3+}$ oxidation state in the nonpolar phase leads to magnetic activity [29], consistent with empirical trends observed in transition metal oxides. Specifically, the magnetic ground state of nonpolar TiGeTe$_3$ was found to be a zigzag-type antiferromagnet (Z-AFM). Importantly, both the ferroelectric and nonpolar Z-AFM phases are dynamically stable and insulating. Furthermore, the DFT calculations revealed that these two ferroic phases are separated by an energy barrier (Fig. 2d), confirming their thermodynamic stability. The Z-AFM phase represents the global ground state, while the ferroelectric phase is metastable (Fig. 2d). From the electronic charge distribution in the top valence bands of TiGeTe$_3$, the $4s^2$ lone pairs of Ge ions in the ferroelectric phase appear as unilateral "hats", indicating nonbonding character. In contrast, the Z-AFM phase exhibits covalent bonding between Ge-Ge pairs (Fig. 2e) This difference in bonding configuration may lead to a change in the valence state of the Ti ions (i.e., Ti$^{4+}$ *vs.* Ti$^{3+}$). Then this variable valence of Ti was theoretically in TiGeSe$_3$ monolayer: Ti$^{4+}$ ($3d^0$) in the polar phase (Fig. 2f) *vs* Ti$^{3+}$ ($3d^1$) in the nonpolar phase (Fig. 2g) [30].

Based on Landau free energy fitting, we identified that TiGeSe$_3$ satisfies the autferroic condition $c_p<c_m<c$ [30], indicating that the magnetic phase is the ground state and the ferroelectric phase is metastable. Further DFT calculations support the metastability of the ferroelectric phase. In particular, the ferroelectric state was found to be robust against small magnetic perturbations, confirming that it does not correspond to an energy saddle point. These results suggest that TiGeSe$_3$ monolayer is also a possible candidate for autferroicity. Furthermore, the transition from the autferroic state to the single magnetic state in TiGeSe$_3$ was predicted to occur at ~63 K (Fig. 2h). The seesaw-type ME coupling in the TiGeSe$_3$ monolayer arises from the intrinsic mutual exclusivity between the ferroelectric and magnetic sublattices, which is fundamentally driven by strong spin-phonon coupling.

In contrast, TiSnSe$_3$ was predicted to be a single-ferroic material [30]. While the TiSnSe$_3$ monolayer exhibits structural and ferroic behavior similar to TiGeSe$_3$, the key difference lies in the relative stability of its ferroic phases: in this case, the ferroelectric phase is energetically favored over the magnetic state. According to the Landau model fitting, TiSnSe$_3$ satisfies the condition $c_m<c<c_p$, indicating that the ferroelectric phase is the ground state, whereas the magnetic state corresponds to an energy saddle point. This conclusion was further supported by DFT calculations, which identified the absence of a stable magnetic phase and, consequently, the improbability of autferroic behavior in TiSnSe$_3$ (Fig. 2i). This trend is physically reasonable. First, the larger radius of Sn ion tends to expand the lattice, which empirically favors ferroelectric distortions by reducing steric hindrance. Second, the more spatially extended $5s^2$ lone pairs of Sn can further enhance ferroelectricity through increased stereochemical activity.

Beyond the TMTC family, competition and transitions between ferroelectric and magnetic orders have also been experimentally observed in the solid solution series [1-$x$](Ca$_{0.6}$Sr$_{0.4}$)$_{1.15}$Tb$_{1.85}$Fe$_2$O$_7$-[$x$]Ca$_3$Ti$_2$O$_7$ [36], although the concept of autferroicity had not been introduced at the time. As such, the ferroic behavior in this system warrants examination in the context of autferroicity in future studies. Moreover, Landau theory further suggests that autferroicity may emerge at finite temperatures even in single-ferroic materials (Fig. 1f), provided that the ME coupling strength is sufficiently strong. This condition offers a promising criterion for identifying additional autferroic candidates.

**Potential applications for autferroicity**

The emergence of autferroicity as a new ferroic state presents exciting opportunities for functional materials and device innovation. The conventional multiferroics often suffer from weak ME coupling in those type-I multiferroics or weaker ferroelectricity in those type-II multiferroics. However, autferroics leverage the mutual exclusivity between these two ferroic orders to enable field-selective switching, robust bistability, and intrinsically strong ME responses. These characteristics offer a promising foundation for diverse applications.

In particular, autferroics may serve as a platform for nonvolatile memory devices, where logic states are encoded not in the orientation of a single order parameter, but in the identity of the active ferroic phase, either magnetic or ferroelectric state. This switching scheme offers several advantages, including high read/write contrast, minimal crosstalk, and potentially low power consumption, owing to the sharp, field-driven transitions between bistable ferroic states. In contrast, conventional multiferroic memory devices require robust primary ferroic order parameters to achieve large ferroic-direction-dependent resistance [37,38], i.e., electroresistance or magnetoresistance, as well as strong ME coupling. However, as discussed above, these requirements are difficult to simultaneously satisfy within the framework of conventional multiferroics.

Beyond binary memory applications, the bistable energy landscape of autferroics, energetically separated ferroic phases, also lends itself to true random number generation (TRNG) [39,40]. The strong and repulsive ME coupling in autferroics reshapes the energy landscape along the switching pathway (Fig. 1c), effectively providing magnetic (or ferroelectric) fluctuations with a "detour" route characterized by a reduced energy barrier. This alternative path facilitates higher switching frequencies and, consequently, an increased rate of TRNG generation, even in the presence of a shallow metastable intermediate state. This characteristic makes autferroic materials strong candidates for hardware-level TRNGs, which are essential for secure communications and probabilistic computing.

From the experimental perspective, the practical realization of autferroic devices will require significant advances in material synthesis, precise phase control, and interface engineering to reliably access the desired autferroic regimes under ambient conditions.

Nevertheless, the foundational concept of exploiting ferroic exclusivity, rather than coexistence, marks a paradigm shift in the design paradigm of functional ferroic materials. Autferroics open new avenues for the development of multifunctional devices beyond the limits of conventional multiferroics.